\documentclass[]{raa}            
\usepackage{graphicx,times}
\usepackage{natbib}

\providecommand{\bm}[1]{\mbox{\boldmath$#1$\unboldmath}}

\begin{document}

   \title{Observation of Interplanetary Scintillation with Single-Station Mode at Urumqi
}

 \volnopage{ {\bf 2009} Vol.\ {\bf 9} No. {\bf XX}, 000--000}
   \setcounter{page}{1}

   \author{Lijia Liu
      \inst{1}
   \and Xizhen Zhang
      \inst{1}
   \and Jianbin Li
      \inst{1}
   \and P.K. Manoharan
      \inst{2}
   \and Zhiyong Liu
     \inst{3}
   \and Bo Peng
      \inst{1}
   }

   \institute{National Astronomical Observatories, Chinese Academy of Sciences,
             Beijing 100012, China; {\it liulijiaredstar@yahoo.com.cn, lijb@bao.ac.cn, zxz@bao.ac.cn, pb@bao.ac.cn}\\
        \and
            {Radio Astronomy Center,TIFR-NCRA,P.O. Box 8,Ooty 643 001,India; }mano@ncra.tifr.res.in
        \and
             {Urumqi Observatory, NAOC
No.150 South Science Road Urumqi, Xinjiang 830011, China;
}liuzhy@uao.ac.cn
\vs \no
}

\abstract{ The Sun affects the Earth's physical phenomena in
multiple ways, in particular the material in interplanetary space
comes from coronal expansion in the form of inhomogeneous plasma
flow (solar wind), which is the primary source of the interplanetary
medium. Ground-based Interplanetary Scintillation (IPS) observations
are an important and effective method for measuring solar wind speed
and the structures of small diameter radio sources. We discuss one
mode of ground-based single-station observations: Single-Station
Single-Frequency (SSSF) mode. To realize the SSSF mode, a new system
has been established at Urumqi Astronomical Observatory (UAO),
China, and a series of experimental observations were carried out
successfully from May to December, 2008.
\keywords{IPS-SSSF-Observation } }

   \authorrunning{Lijia Liu, Xizhen Zhang,Jianbin Li, P.K. Manoharan, Zhiyong Liu \& Bo Peng }            
   \titlerunning{Observation of Interplanetary Scintillation with Single-Station Mode at Urumqi}  
   \maketitle


%
%
\section{Introduction}           
\label{sect:intro}

Radiation from a distant compact radio source is scattered by the
density irregularities in the solar wind plasma and produces a
random diffraction pattern on the ground.  The motion of these
irregularities converts this pattern into temporal intensity
fluctuations which are observed as interplanetary scintillation
(IPS). IPS observations with ground-based telescopes can estimate
 the solar wind velocity and also the structures of the distant compact
radio sources (Hewish \& Symonds 1969; Armstrong \& Coles 1972).
This kind of measurement, though indirect, can give information on
the solar wind out of the ecliptic plane and close to the Sun, where
direct spacecraft measurements are not possible (e.g. Ma 1993). Here
we concentrate on extracting information on solar wind speed from
IPS observations with ground-based single telescope.

There are two modes to observe the IPS phenomenon with a
single-station: Single-Station Single-Frequency (SSSF) and
Single-Station Dual-Frequency (SSDF). Since the discovery of the IPS
phenomenon (Hewish et.al. 1964), many countries began doing IPS
observations with the single-station method, i.e.
 Cambridge telescope in Britain (Pruvis et al. 1987),
Ooty radio telescope in India (Swarup et al. 1971), Puschino
observatory in Russia (Vitkevich et al. 1976), one can use the power
spectral fitting method to obtain the solar wind speed. With the
multi-station method as used in Japan, which is a three-station
system (Kojima et al. 1995), one can measure the projected solar
wind speed directly. China began IPS studies from the 1990s with the
phased array mode of the Miyun Synthesis Radio Telescope (MSRT) at
232 MHz. Located at Miyun observatory (Wang 1990) in Beijing, it
used the SSSF mode (Wu \&  Zhang 2001). Recently a new IPS
observation system using the 50 m parabolic radio telescope, which
is based on the SSDF mode at S/X and UHF bands, is under
construction to serve the National Meridian Project of China.

In this paper, the theory of IPS, the technique, and the numerical
simulation method of SSSF mode are introduced in section 2. The
observations carried out at the UAO are reported in section 3.
Finally conclusions are drawn in Section 4.


\section{THEORY}
\subsection{The Theory of IPS}
\label{sect:Obs}

Interplanetary scintillation is the name given to the intensity
fluctuations of small diameter radio sources which are caused by
density inhomogeneities in the solar wind.  Figure 1 shows the
geometry of IPS.The distance between the Sun and $Q$ is $r$, and
 $r=sin(\varepsilon)AU$ .

\begin{figure}[ht]
\center{\includegraphics[width=0.5\columnwidth]{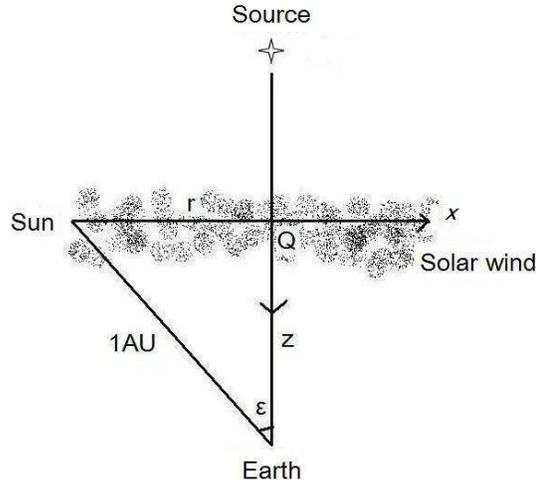}
\caption{IPS Geometry: \textit{z}-axis is along the line-of-sight,
\textit{x}-axis is in the direction perpendicular to the
\textit{z}-axis pointing away from the Sun, and \textit{y}-axis is
normal to the paper. $Q$ is the point closest to the Sun along the
line-of-sight, and $\varepsilon$ is the elongation angle,
Sun-Earth-source. $r$ is the distance between the Sun and $Q$. $Z$
is the distance between $Q$ and the Earth. }}\label{1}
\end{figure}

The degree of scintillation is characterized by the scintillation
index $m$ (Cohen et al. 1967), which increases with decreasing
distance $r$ until it reaches a maximum $m_{max}$ at $r_{min}$.
$r_{min}$ changes with frequency, $r_{min}\approx0.2 AU$ for meter
wavelength (Manoharan, 1993).
\begin{eqnarray}\label{1}
m=\frac{\sqrt{\sigma^{2}_{on}-\sigma^{2}_{off}}}{C_{on}-C_{off}}
\end{eqnarray}
where $C_{on}$ ($C_{off}$) is the average intensity of the on-source
(off-source) signal, and $\sigma^{2}_{on}$ ($\sigma^{2}_{off}$) is
the square of rms of intensity scintillation. Here on-source case
refers the telescope pointing at the radio source, and off-source
the telescope pointing at the background sky away from the source,
being in the opposite pitching direction to where the radio source
moves. IPS is strongest in the region nearest the Sun, where we have
the "strong scintillation region". In most of interplanetary space
IPS is weak, which is called the "weak scintillation region" (Zhang
2006). In the weak scintillation region, $m^{2}\ll1$.  The weak
scintillation region is observed at $r > r_{min}$ and strong
scintillation is observed at $r < r_{min}$. Previous studies show
that the statistics of the scintillation are simply related to those
of the turbulent interplanetary medium by a linear relationship, if
the scintillation is weak (Coles \& Harmon 1978). In the "strong
scintillation region", however, the relationship is not straight
forward, and the present study always deals with the weak
scintillation case. The distance regime for the weak and strong
regions relates to observing frequencies. Table 1 shows the
relationship between the frequency and distance regimes (Zhang
2006).

\begin{table}[ht]
\center{\includegraphics[width=0.8\columnwidth]{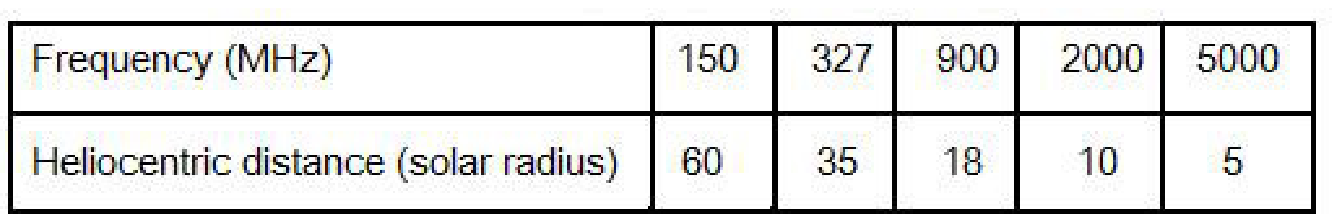}
\caption{Regions of strong and weak scintillation at different
frequencies. Taking 327 MHz for example, the regimes for the strong
and weak regions are within 35 solar radius and beyond
respectively.}}\label{1}
\end{table}

In the weak scintillation region, where the radio wave can be
treated as a plane-wave, and the Born approximation is applicable
(e.g. Walker et al. 2004), the interplanetary medium can be
considered to be made up of many thin layers perpendicular to the
line-of-sight. When the radio wave passes through these layers, only
 the phase of the radio wave changes, while the amplitude of the
radio wave stays the same. This is called the "thin screen
approximation", which is commonly used in the study of the
ionosphere, interplanetary medium, and interstellar medium.

\subsection{SSSF Mode}
SSSF refers to observing the IPS with a single station at a single
frequency. There are two methods to obtain the solar wind speed from
SSSF mode observed spectra: the spectral multi-parameter
model-fitting, and the characteristic frequencies methods.  The
former can measure the speed by adjusting the main parameters of the
solar wind to fit the observed scintillation power spectra.  The
parameters are: $\alpha$-power law index of the spatial spectrum of
electron density, AR-axial ratio of solar wind irregularities, and
$V$-solar wind speed. The latter can be determined by calculating
two characteristic frequencies of the spectra: the Fresnel knee
frequency $f_{F}$, and $f_{min}$ the first minimum of the spectra.
Then the solar wind speed can be calculated by either of the
formulas shown below (Scott et al. 1983)

\begin{eqnarray}\label{2}
V=f_{F}\sqrt{Z\pi\lambda}
\end{eqnarray}

\begin{eqnarray}\label{3}
V=f_{min}\sqrt{Z\lambda}
\end{eqnarray}

where $\lambda$ is the observing wavelength, $Z$ is the distance
between $Q$ and the Earth as shown in Fig. 1.  According to weak
scintillation theory and thin screen approximation theory, in the
weak scintillation region, the observed scintillation can be
regarded as the sum of contributions from all the thin layers. For a
layer of thickness $dZ$, the distance from the layer to the Earth is
$Z$, $V_{x}(\textit{z})$ is the solar wind velocity projected on to
the plane perpendicular to the direction $\textit{z}$, and the
spectra observed at the Earth should be (e.g. Scott et al. 1983; Ye
\& Qiu 1996):

\begin{eqnarray}\label{4}
M_{s}(f,Z)dZ&=&\frac{2\pi f(\lambda r_{e})^{2}}{V_{x}(\textit{z})}
\int^{\infty}_{-\infty}\Phi_{ne}(k_{\textit{x}},k_{\textit{y}},k_{\textit{\textit{z}}}=0,Z)\times
F_{d}F_{s}dk_{\textit{y}}dZ
\end{eqnarray}
where,
\begin{eqnarray}\label{5}
F_{d}=4\sin^{2}[\frac{(k_{\textit{x}}^{2}+k_{\textit{y}}^{2})\lambda
Z}{4\pi}]
\end{eqnarray}

\begin{eqnarray}\label{6}
F_{s}=\exp[-(k_{\textit{x}}^{2}+k_{\textit{y}}^{2})Z^{2}\theta_{0}^{2}]
\end{eqnarray}

Here $F_{d}$ and $F_{s}$ are the Fresnel propagation filter
parameter and the squared modulus of the radio source visibility. We
assume that the brightness of the radio source has a
symmetrical-Gaussian distribution, i.e.
$B(\theta)=\exp[\frac{-(\theta /\theta_{0})^{2}}{2}]$. $\Theta$ is
the full width at half maximum of the source, so we have the angular
diameter of the scintillating source $\theta_{0}=\Theta/2.35$
(Manoharan \& Ananthakrishnan 1990). $\Phi_{ne}$ is the electron
density power spectrum at distance $z$,

\begin{eqnarray}\label{7}
\Phi_{ne}(k_{\textit{x}},k_{\textit{y}},k_{\textit{z}}=0,Z)=Tr^{-4}[k_{\textit{x}}^{2}+(k_{\textit{y}}/AR)^2]^{-\alpha
/2}
\end{eqnarray}
where the amplitude of fluctuations in the electron density $T$ is a
constant. The spectrum obtained from the Earth should be the sum of
Eq. (4).
\begin{eqnarray}\label{8}
M_{i}(f)=\int_{0}^{\infty}M_{s}(f,Z)dZ
\end{eqnarray}

One can see that $M_{i}(f)$ depends on the solar wind parameters:
axis ratio $AR$, power law index $\alpha$, and the solar wind speed
$V$. Previous studies show that, when other parameters fixed, $AR$
mostly affects the low frequency part of the power spectra; when
$AR$ increases, the low frequency part becomes steeper, but the high
frequency part changes slightly. $\alpha$ mostly affects the high
frequency part of the spectra; when $\alpha$ increases, the high
frequency part attenuates quickly, but the low frequency part
changes inevidently. The solar wind speed $V$ mostly affects the
Fresnel knee $f_{F}$ and the first minimum of the spectra $f_{min}$;
the two frequencies both become larger when $V$ increases. Taking
appropriate values to fit the observed spectra, one can obtain the
parameters of the observational data. Firstly one fits the fresnel
knee according to $f_{F}$ and $f_{min}$, then fits the attenuated
high part and the flat low frequency part. (Ye \& Qiu 1996)

Fig.2 is an example of the SSSF mode. Where the parameters taken are
$\lambda=92 cm$, $\alpha=3.5$, $AR=2.0$, $V=600 km/s$,
$\theta_{0}=0.02''$. One can get $f_{F}=1.05 Hz$, then from Eq.(2)
we can obtain the solar wind speed, which is 598.7 km/s, which fits
to the simulated value well.

\begin{figure}[ht]
\center{\includegraphics[width=0.5\columnwidth]{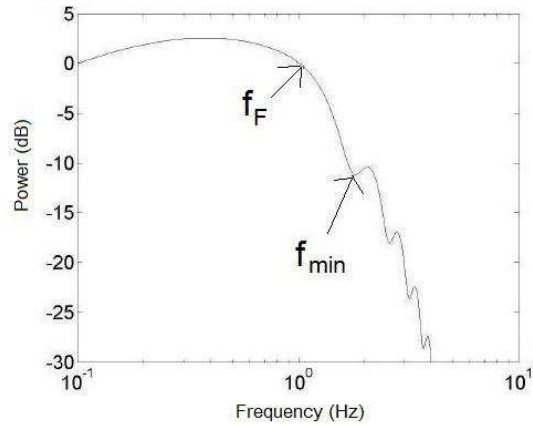}
\caption{Simulation result of SSSF mode, $\lambda=92 cm$,
$\alpha=3.5$, $AR=2.0$, $V=600 km/s$, $f_{F}=1.05 Hz$}}\label{2}
\end{figure}

\section{OBSERVATIONS}
\subsection{Instrument setup}
The IPS experimental observations were performed from May to
December 2008 with the 25 m radio telescope at UAO, China, in SSSF
mode. The UAO radio telescope is located to the south of Urumqi
city, with 87 deg longitude and 43 deg North latitude. Table 2 shows
general information for the 25 m radio telescope receivers currently
available at UAO.

\begin{table}[ht]
\center{\includegraphics[width=0.65\columnwidth]{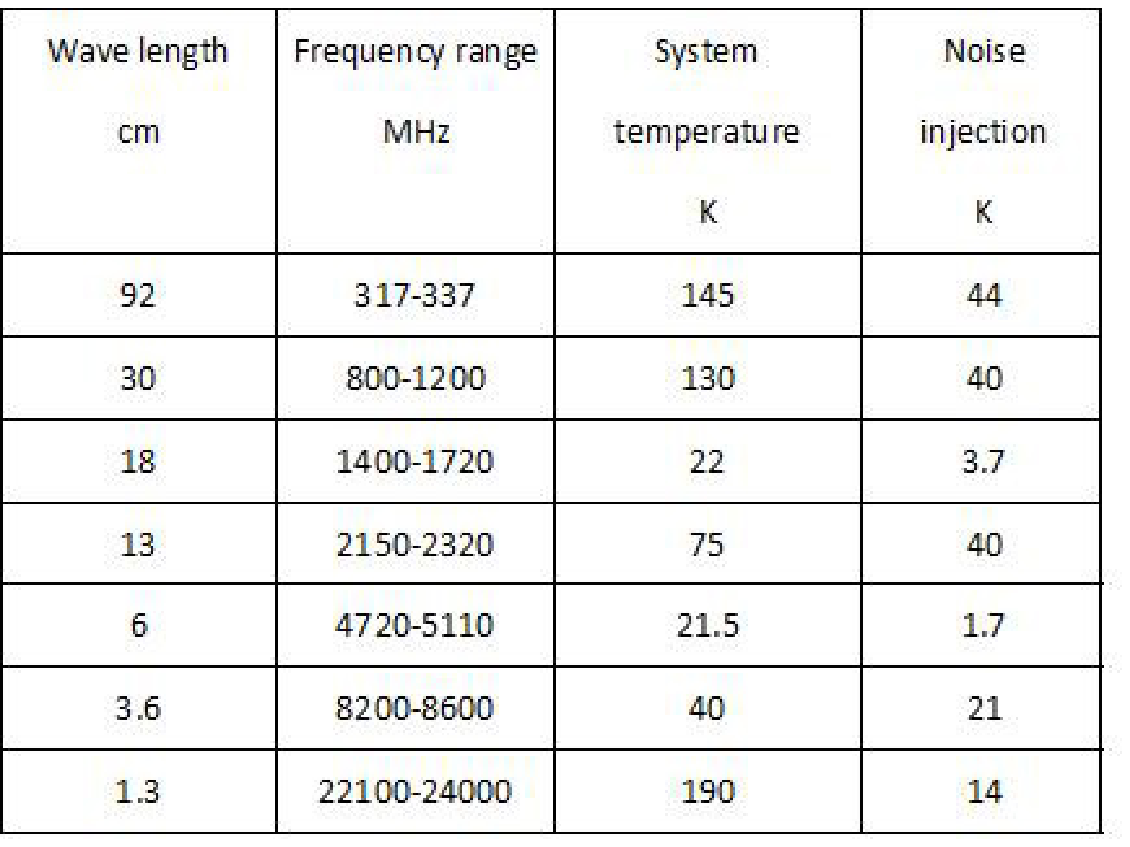}
\caption{Information on the current receivers at UAO, where columns
1-4 give wavelength in cm, frequency range in MHz, system
temperature in K, and noise injection in K.}}\label{2}
\end{table}

The 6 cm and 18 cm bands have dual-polarization cooled receivers,
while the 3.6 cm and 13 cm bands have single polarization cooled
receivers. Table 3 summarizes some information on the IPS
observations we carried out.  Compact, strong radio sources selected
for observation are listed in table 4.

\begin{table}[ht]
\center{\includegraphics[width=0.65\columnwidth]{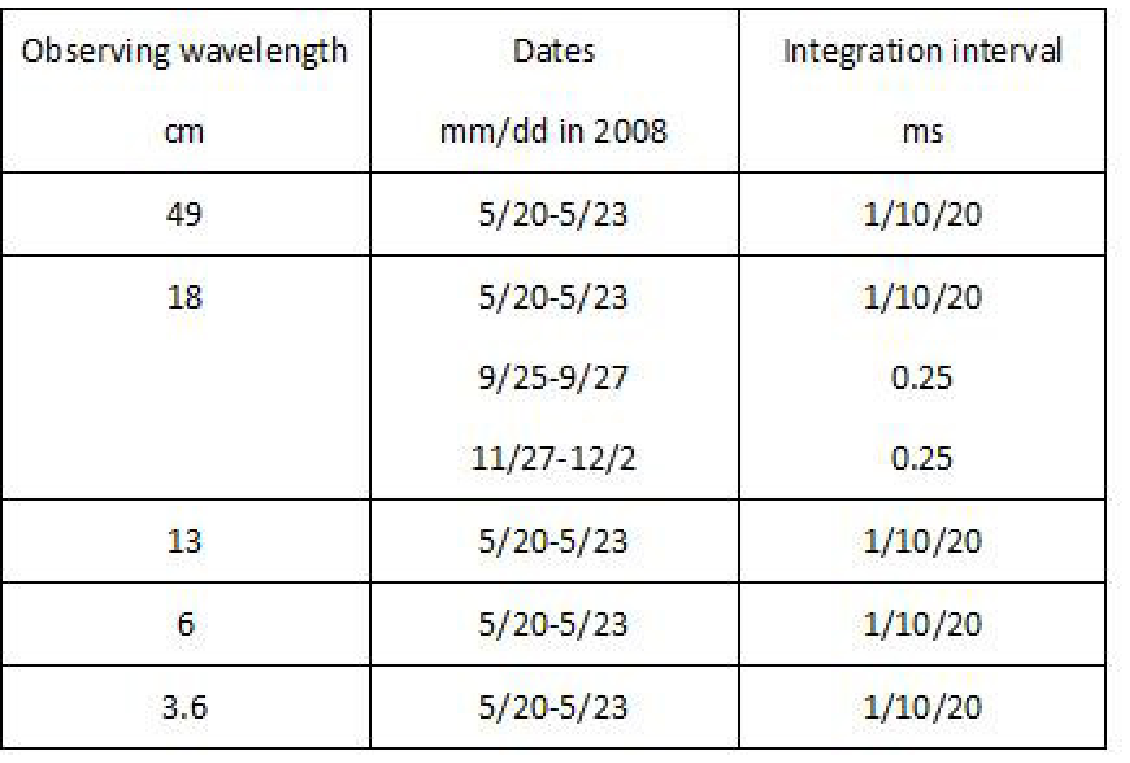}
\caption{Key parameters of IPS observations with 25 m radio
telescope performed in 2008 at UAO, where columns 1-3 give observing
wavelength in cm, observing date in mm/dd, and integration interval
in ms.}}\label{3}
\end{table}

\begin{table}[ht]
\center
 {\includegraphics[width=0.65\columnwidth]{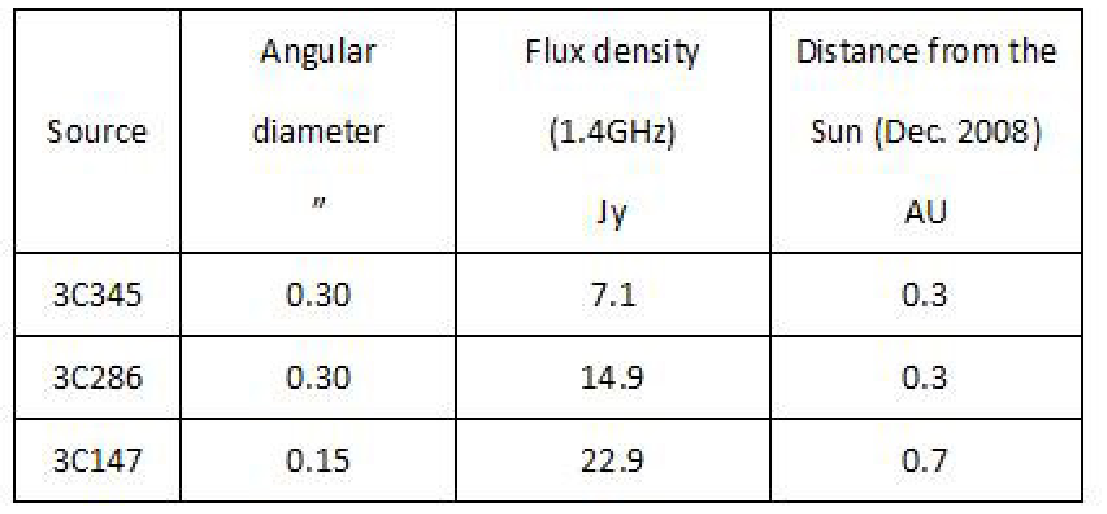}
\caption{Details of the observed sources, where columns 1-4 give the
source name, angular diameter of the source in arc-second, the flux
density at 1.4GHz in Jy and the distance from the Sun in
AU.}}\label{4}
\end{table}

Test observations were carried out at 49 cm, 18 cm, 13/3.6 cm and 6
cm. Each time the observation was performed on-source for 10
minutes, and off-source for 5 minutes.  The integration intervals
tried were 1 ms, 5 ms and 10 ms.

According to the characteristics of the IPS phenomenon and
synchrotron radiation of radio sources, it would be easier to detect
IPS at lower observing frequencies.  While the radio environment at
UAO are not good at the 92 cm and 49 cm, so they are seldom used.
Consequently we concluded that the 18 cm band is the only window
suitable for catching IPS at UAO.

After a series of experiments, the 18 cm dual-polarization receiver
at UAO was chosen for the observations, and a data
acquisition/receiving system was also established. The data sampling
rate is adjustable with 8-bit quantification rate. Being a real-time
display system, data quality can be monitored during the
observation, so parameters like gain or target source can be
adjusted immediately. In order to minimize the RFI (radio frequency
interference) influence in the observing window, a band-pass filter
was added to the output of IF (intermediate-frequency) of the 18 cm
receiver. Figs.3 and 4 show a characteristic spectrum of the 18 cm
receiver before and after the filter was added.

\begin{figure}[ht]
\center {\includegraphics[width=0.5\columnwidth]{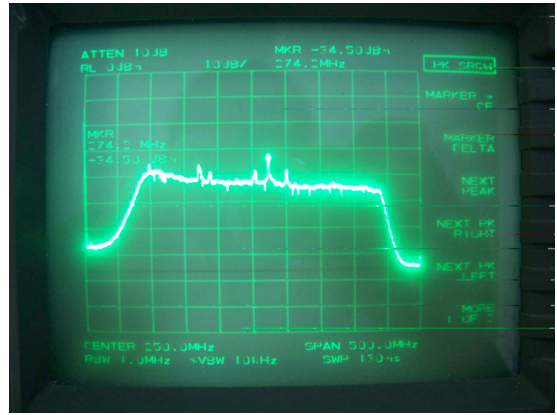}
{\caption{Characteristic spectrum of 18cm receiver at UAO before
filtering }}}\label{3}
\end{figure}

The entire bandwidth at UAO is 500MHz, with some interference in the
band as shown in Fig.3. The central frequency of the filter was set
to 420MHz, and the 3dB bandwidth was 100MHz, with an insertion loss
of 3dB.  It can be seen from Fig.4 that the filter works well, the
interference in this band has been effectively filtered out. The
band selected was the part that with the lowest interference of the
whole band. The filter introduces some loss, so we added an
amplifier before the radiometer but after the filter.

\begin{figure}[ht]
\center {\includegraphics[width=0.5\columnwidth]{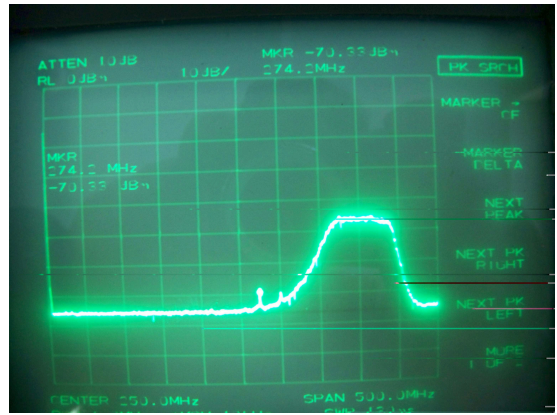}
{\caption{Characteristic spectrum of 18 cm receiver at UAO  after
filtering}}}\label{4}
\end{figure}


\begin{figure}[ht]
\center{\includegraphics[width=0.45\columnwidth]{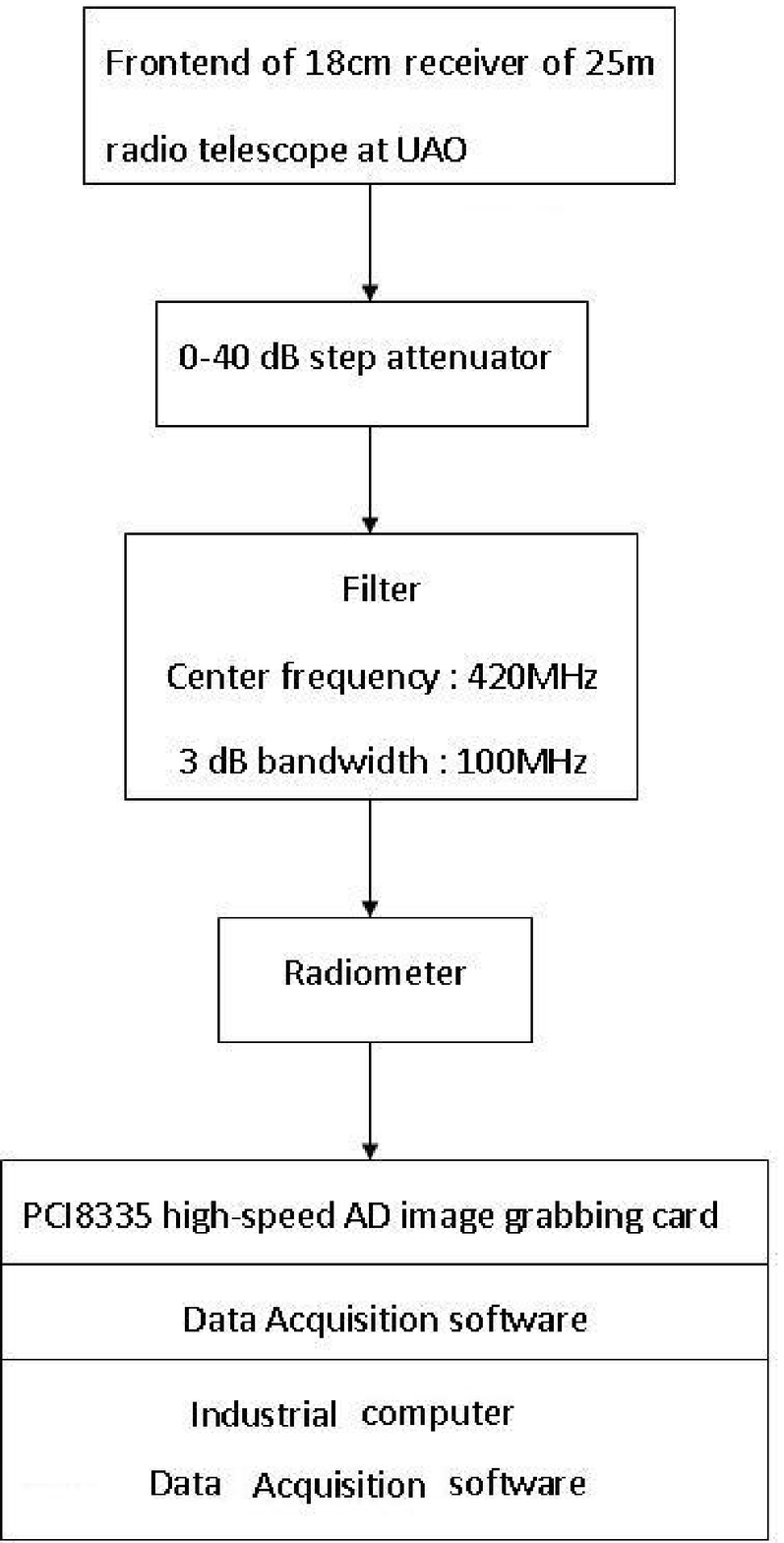}
\caption{Flowchart of data acquisition instrument at UAO for IPS
observations}}\label{5}
\end{figure}

Fig. 5 is a flowchart of the data acquisition instrument.  There is
a 0-40 dB step attenuator after the frontend of the 18 cm receiver,
with the attenuation step of 1 dB.  A PCI8335 high-speed AD image
acquisition card was added to our industrial computer, with an input
voltage range 0-5V.  The AD precision of this card is 16 Bit, with
maximum of sampling rate of 250 kHz, and the buffer (FIFO: first in
first out) is 8 Kbytes.  The radio-meter has two channel outputs.
The band of channel A is 5-500 MHz, and the band of channel B is
400-950 MHz.   The input power for the two channels is the same: -20
dBm to -60d Bm, and the output voltage range of the radiometer is
0-5 V.   Channel A was used during our observation. Raw data,
together with information on the target source like observing time,
source coordinates etc., are recorded by the data acquisition
software.

During the observations each time the on-source observations were 10
to 15 minutes, and the off-source observations were 5 minutes. In
view of the different distances and orientations with respect to the
sun, we observed different sources at different times.  The total
observing time each day was about 2-3 hrs.

\subsection{Data analysis}

In order to eliminate the interference, besides the hardware method
(adding a filter), a software solution has also been developed.
Figure 6 is the flowchart of the data analysis. First, the raw data
observed are played back on the screen to identify the parts with
lower noise and one subtracts the noise using software, i.e. the
slowly changing component is subtracted from the raw data, and
assigned to DATA1. DATA1 is then compared with 3 times the rms error
of a long span of data to eliminate wild points. Points with
absolute values higher than the 3 rms are omitted and replaced by
the average value of the preceding and following data points, then
this data is assigned to DATA2. The original integration interval of
an observation being 0.25 ms, we take the average of four contiguous
points to form a 1 ms integration dataset, which means the 10 ms of
data are obtained by averaging 40 contiguous points. This data is
assigned to DATA3.

\begin{figure}[ht]
\center{\includegraphics[width=0.45\columnwidth]{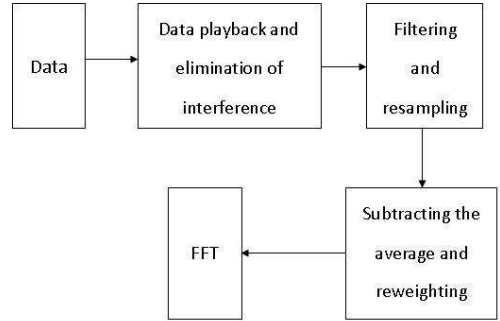}
\caption{Flowchart of data reduction for IPS observations}}\label{6}
\end{figure}

In the filtering and re-sampling step, DATA3 is convolved with a
rectangular window of suitable width corresponding to the
re-sampling rate. The time series is then broken into blocks of
length 8192 samples (for 1 ms data approximately 10 s long), and the
mean value of each block is subtracted and the block is then
multiplied by a triangular weighting function, which is unity at the
center and falls to zero at both ends.  The result is then
transformed by Fourier transformation (FT) to obtain the power
spectrum.

\subsection{Observational results}
A series experimental IPS observations were made at UAO.  Fig. 7
shows raw data obtained on Nov. 27. The pointed source was 2MASX
J18141308-1755351. Its flux density at 1.4 GHz is 5.39 Jy, and its
projected distance from sun was 0.23 AU.

\begin{figure}[ht]
\flushleft\ \ \ \ \ \ \ \ \ \ \ \ \ \ \ \ \ \ \ \ \ \ \ \ \ \ \ \ \
\ \ \ \ \ \ \ \ \ \ \ \ \
{\includegraphics[width=0.5\columnwidth]{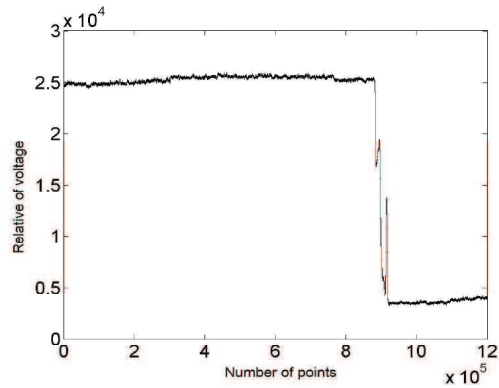} {\caption{Raw
dataset display for 2MASX J18141308-1755351, observed on Nov.27,
2008. The \textit{x}-axis is the number of points, the
\textit{y}-axis is the relative of voltage, each point taken with
0.25 ms sampling rate}}}\label{7}
\end{figure}

One can see that the on-source part and off-source part are
obviously identified.  The fluctuation of the two parts are almost
the same, which indicates the IPS phenomenon at the time was weak,
which is identical to the power spectrum in Fig. 8.  It is clear
that the Fresnel knee $f_{F}$ and the first minimum frequency
$f_{min}$ are difficult to identify, indicating that there was
little scintillation.

\begin{figure}[ht]

\flushleft\ \ \ \ \ \ \ \ \ \ \ \ \ \ \ \ \ \ \ \ \ \ \ \ \ \ \ \ \
\ \ \ \ \ \ \ \ \ \ \ \
{\includegraphics[width=0.5\columnwidth]{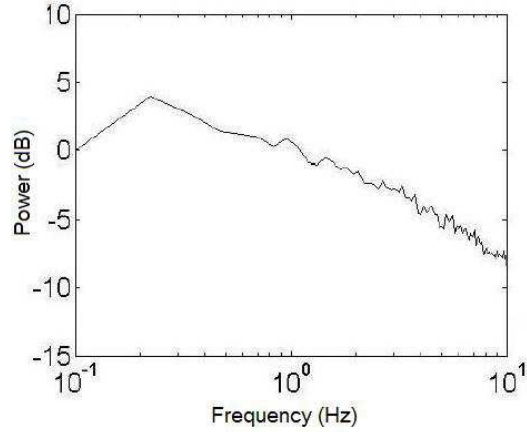} \caption{Power
spectrum result of the target source 2MASX J18141308-1755351 at 18
cm, observed on Nov.27, 2008. The \textit{x}-axis is power, and the
\textit{y}-axis is frequency }}\label{8}
\end{figure}

Fig. 9 shows a model-fit result of the data taken on Dec. 1, at a
wavelength of 18 cm, with an integration interval of 1 ms. The
scintillation index is in the range 0.6 to 0.7 (There are some
interference in the off source part).  The model-fit method is the
same as that of SSSF simulation. According to equation (4) to (8),
the best parameters can be obtained by fitting with the observing
spectra. The target source was 3C345, with a flux density at 1.4 GHz
of 7.1 Jy. The solid line shows the observed power spectrum, and the
dashed line is the result of parametric model-fitting, with the
fitting parameters: $AR=1.2$, $\alpha=3.1$, $V=400 km/s$.  According
to OMNI data base, the solar wind speed the whole day ranged between
300 to 400 km/s, which is in agreement with the model-fitted value.

\begin{figure}[ht]
\flushleft\ \ \ \ \ \ \ \ \ \ \ \ \ \ \ \ \ \ \ \ \ \ \ \ \ \ \ \ \
\ \ \ \ \ \ \ \ \ \ \
{\includegraphics[width=0.55\columnwidth]{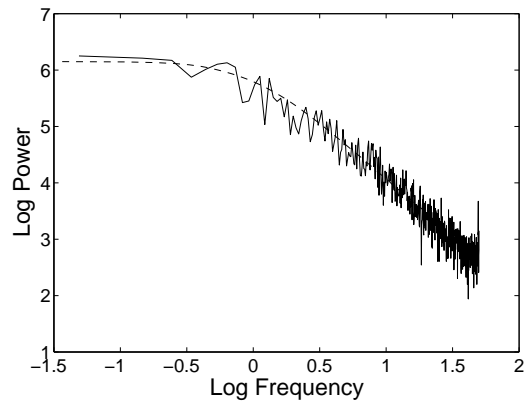} \caption{Model
fitting result on 3C345, observed on Dec.1, 2008 at 18 cm. The solid
line shows the observed spectrum of the data, the dashed line is the
result of parametric model-fitting.}}\label{9}
\end{figure}

\section{CONCLUSIONS}

The SSSF mode IPS observations have been studied by quite a number
of pioneers (Manoharan et al. 1994). Its instrument, data
acquisition, and data reduction are simple. For this mode, high
signal-to-noise ratio (at least 25dB) data are needed (Tokumaru et
al. 1994). When AR increases, $f_{F}$ becomes ambiguous, and
$f_{min}$ is easily affected by noise, AR and $\varepsilon$.  The
fitting accuracy is affected by variations in the solar wind
parameters, making it hard to calculate the solar wind speed
accurately.

Compared with the SSSF mode, the SSDF technique gives the solar wind
speed via the first zero point of the cross correlation spectrum,
and $f_{zero}$ is most apparently affected by the velocity of the
solar wind rather than other parameters (i.e.Zhang 2007).  It has
the advantages of higher accuracy on the measurement of solar wind
speed and higher stability against the wide variations in solar wind
parameters.  But it introduces more complexity in the observing
instrument and data taking system, it is not used as widely as the
SSSF mode.

The new system that is under construction at Miyun station near
Beijing, China, with the 50 m radio telescope, adopted the SSDF mode
to do the IPS observations.  There are some lessons to be learned
from the observations with the UAO 25 m radio telescope, such as the
integration time of the receiver system should be sufficiently short
since the IPS phenomenon varies rapidly. This implies that the
effective receiving area of an IPS antenna should be large enough to
ensure that the system has a high instantaneous sensitivity and its
band-width should be well-matched to the system time resolution.  A
bandpass filter and low noise amplifiers (LNA) would be needed to
reduce the system noise level.

\section{ACKNOWLEDGEMENTS}
The authors thank all the staff of Urumqi Astronomical Observatory,
National Astronomical Observatories, Chinese Academy of Sciences,
especially Yi Aili, Yu Aili, N. Wang, X. Liu, H.G. Song, for their
help during the observations. We are also grateful to T.Y. Piao and
Y.H. Qiu, H.S. Chen, W.J. Han, C.M. Zhang, Y.J. Zheng, for their
encouragement and helpful discussions. This work has been supported
by the National Meridian Project(grant no.[2006]2176).

\label{lastpage}

\end{document}